\documentclass[a4paper,twocolumn,11pt, unpublished]{quantumarticle} 
\pdfoutput=1
\usepackage[utf8]{inputenc}
\usepackage[english]{babel}
\usepackage[T1]{fontenc}
\usepackage{float}
\usepackage{amsmath, amsfonts,physics}
\usepackage{booktabs} 
\usepackage[citecolor=blue, linkcolor=teal, urlcolor=violet]{hyperref}
\usepackage{xcolor}

\usepackage{graphicx}
\usepackage{multicol}
\usepackage[ruled, lined, linesnumbered, commentsnumbered, longend]{algorithm2e}

\newtheorem{theorem}{Theorem}
\newtheorem{dfn}{Definition}

\begin{document}

\title{Graph test of controllability in qubit arrays: A systematic way to determine the minimum number of external controls}

\author[1]{Fernando Gago-Encinas}
\author[1]{Monika Leibscher}
\author[1]{Christiane P. Koch}
\affil[1]{Fachbereich Physik and Dahlem Center for Complex Quantum Systems, Freie Universit\"{a}t Berlin, Arnimallee 14, 14195 Berlin, Germany}
\maketitle

\begin{abstract}
The ability to implement any desired quantum logic gate on a quantum processing unit is equivalent to evolution-operator controllability of the qubits. Conversely, controllability analysis can be used to minimize the resources, i.e., the number of external controls and qubit-qubit couplings, required for universal quantum computing. Standard controllability analysis, consisting in the construction of the dynamical Lie algebra, is, however, impractical already for a comparatively small number of qubits. Here, we show how to leverage an alternative approach, based on a graph representation of the Hamiltonian, to determine controllability of arrays of coupled qubits. We provide a complete computational framework and exemplify it for arrays of five qubits, inspired by the \textit{ibmq\_quito} architecture.
We find that the number of controls can be reduced from
five to one for complex qubit-qubit couplings and to two for standard qubit-qubit couplings.
\end{abstract}


\section{Introduction}

Universal quantum computing requires evolution-operator controllability on the quantum processing units in order to perform every possible quantum logic gate. One way to achieve this, pursued for example in most superconducting qubit architectures, is to couple each qubit to at least one other qubit and drive all qubits locally. This approach becomes impractical for larger qubit arrays due to increasing requirements on physical space for control lines and on calibration time for control pulses. One may wonder whether a number of local drives smaller than the number of qubits would already be sufficient. If so, this would suggest the possibility of more resource-efficient architectures than currently in use.

Here, we show that controllability analysis provides a systematic approach to determine the minimum number of local controls for which any desired quantum logic gate can be implemented. We find that indeed the number of local controls can be smaller than the number of qubits. The minimum number of local controls, for a given size of the qubit array, depends on the type of qubit-qubit couplings. To facilitate analysis of medium-sized qubit arrays, we leverage a graph theory-based approach to controllability analysis and investigate five-qubit arrays starting from the \textit{ibmq\_quito} architecture. 

Controllability analysis answers the question, in a yes-no fashion,  which states can be reached by time evolution from a set of initial states~\cite{GlaserEPJD15,KochEPJQT22}. The standard approach consists in determining the rank of the dynamical Lie algebra of the system and comparing it to the dimension of the algebra that generates the unitary group of all time evolutions~\cite{dAlessandro2008}. This approach has been extensively used to prove controllability for finite-dimensional systems with sufficiently small Hilbert space dimension~\cite{SchirmerPRA2001,FuJPhysA2001,AltafiniJMP2002} and can also be employed in subspaces of infinite-dimensional systems, when combined with Galerkin-type approximations \cite{chambrion2009,Boussaid2013,BCCS,leibscher2022}. The focus of these studies has been on single quantum systems. For multi-partite systems, where the Hilbert space dimension scales exponentially in the number of subsystems, evaluating the Lie rank condition quickly becomes challenging. It has thus mainly been used to identify controllable subspaces in systems which as a whole are not controllable~\cite{WangIEEETAC2012,WangPRA2016,ChenPRA2020,albertini2021subspace}. Positive controllability results for qubit arrays, as needed for universal quantum computing, require an alternative approach.

The exponential scaling of Hilbert space dimension with the number of qubits is evidently inherent to gate-based quantum computing and represents a fundamental obstacle to controllability analysis that cannot be overcome with classical computers. But even for qubit arrays which are classically simulable, use of the Lie rank condition is hampered by numerical instabilities which are common  when constructing orthogonal bases in large operator spaces. 
This latter obstacle can be avoided by resorting to graph theory-based methods for controllability analysis~\cite{AlbertiniLinAlg2002,BCCS,boscain2014multi} which have successfully been applied to quantum walks~\cite{godsil2010graph}, quantum networks~\cite{GoklerPRL17}, and quantum rotors~\cite{leibscher2022,pozzoli2022}. The latter are characterized by a highly degenerate spectrum which results in multiple resonant transitions, i.e., transitions with the same resonance frequency that are driven by the same external control. Controllability can then not simply be read off from the connectivity of the graph. This problem is also present for arrays of locally driven coupled qubits but the different spectral structure requires different graphical methods to prove controllability, as we will discuss below.

The manuscript is organized as follows. The basic concepts of controllability analysis are briefly reviewed in \autoref{sec:controllability}. The methodology we suggest to use for controllability analysis of coupled qubit arrays is  presented in \autoref{sec:algorithm} with \autoref{ssec:notation-and-tools} explaining how to deal with resonant transitions, \autoref{ssec:description-algorithms} presenting the actual graph test in the form of a flow chart and three algorithms and \autoref{ssec:simple-examples} illustrating the use of the algorithms on simple two-qubit examples. Our results on five-qubit arrays inspired by the \textit{ibmq\_quito} architecture are presented in \autoref{sec:results}, and \autoref{sec:conclusion} concludes.


\section{Controllability analysis}\label{sec:controllability}
We study quantum systems that couple linearly to controls, such that their Hamiltonian can be expressed as 
\begin{equation}
    \hat{H}(t) = \hat{H}(t; u_1, ... u_m) = \hat{H}_0 + \sum_{j=1}^m u_j (t) \hat{H}_j\,,\label{eqn:controlH}
\end{equation}
where the controls $u_j (t)$ are real-valued functions and $\hat{H}_j$ are the control operators. A state that evolves under \autoref{eqn:controlH}, is given by $\ket{\psi (t)} = \hat{U}(t; u_1, ... u_m) \ket{\psi (0)}$. We consider coherent evolutions because it is the relevant foundation for gate-based quantum information. 

For qubit arrays, the drift $\hat{H}_{0}$ is split into $\hat{H}_0 = \sum_{l=1}^{N_Q} \hat{H}_{0, l} +\hat{H}_c$, representing the independent local Hamiltonians $\hat{H}_{0, l}$ of the $N_Q$ free qubits and the time-independent couplings $\hat{H}_c$ between them. Typically, local controls $\hat{H}_j$ act on single qubits, for example in the form of laser pulses in ion arrays \cite{figgatt2019ion} or varying microwave fields for certain superconducting qubits \cite{krantz2019quantum}. Regardless of the physical implementation of the qubits, we  would like to answer the question whether a system with a set of controls is capable of performing any unitary operation. 

The answer can be found by analysing the controllability; in particular, the evolution-operator controllability of the system \cite{dAlessandro2008}. It is defined as follows:
\begin{dfn} 
A system with controls as defined in \autoref{eqn:controlH} and Hilbert space dimension $n$ is \textbf{evolution-operator controllable} iff for every unitary evolution $\hat{U}_{target} \in U(n)$ there exist a final time $T\geq 0$, a phase angle $\theta \in [0,2\pi)$ and a set of controls $\{u_j\}_{j=1}^m$ such that $\ket{\psi (T)} = e^{i \theta} \hat{U}_{target}\ket{\psi (0)} =  \hat{U}(t; u_1, ... u_m) \ket{\psi (0)}$ for any initial state $\ket{\psi (0)}$ in the system's Hilbert space.  
\end{dfn}
Note that $e^{i \theta}$ is a global phase that does not depend on the choice of the initial state. In other words, a system is evolution-operator controllable if and only if we can always choose controls and a final time to carry out any unitary operation, up to a global phase.  For simplicity, we refer to this property as 'controllable'. 

A widely used method for studying the controllability of a quantum system is to analyze the \textbf{dynamical Lie algebra} \cite{dAlessandro2008, godsil2010graph, schirmer2002identification}. It is generated by the drift and the $m$ control operators of the system, $\mathcal{L} := Lie(i\hat{H}_0, i\hat{H}_1, ..., i\hat{H}_m)$, i.e., it contains the skew-Hermitian operators $i\hat{H}_0, i\hat{H}_1, ..., i\hat{H}_m$, and their (nested) commutators. 
A system with Hilbert space dimension $n$ is controllable if the dimension of the dynamical Lie algebra is $n^2$ or $n^2-1$. In other words, it is controllable if the dimension of the Lie algebra matches either that of $\mathfrak{su}(n)$ (which generates the special unitary group $SU(n)$) or $\mathfrak{u}(n)$ (which generates the unitary group $U(n)$). This is sometimes referred to as the Lie algebra rank condition \cite{dAlessandro2008}. The main difference between generating $U(n)$ or $SU(n)$ is that the former allows the system to perform any unitary evolution including all global phases. Conversely, the latter encompasses all unitary evolutions up to a non-controllable global phase.

Alternatively to constructing the dynamical Lie algebra, a graph test can be used to analyze controllability \cite{boscain2014multi, leibscher2022, boscain2021classical}. The graph encodes the information of the Hamiltonian.
\begin{dfn} 
Given a quantum system evolving under \autoref{eqn:controlH}, we can construct an undirected graph according to the rules: 

\begin{enumerate}
    \item For every eigenstate $\ket{e_k}$ of $\hat{H}_0$, add a vertex to the graph with a corresponding label. 
    \item For every control $\hat{H}_j$ and every nonzero element $\bra{e_i}\hat{H}_j\ket{e_k}$, add an edge between the vertices labelled $\ket{e_i}$ and $\ket{e_k}$. Additionally, add a label to every edge stating which control drives that transition. 
\end{enumerate}

We call this the \textbf{graph of a quantum system}\footnote{Note that given two vertices there can be more than one edge between the two of them (belonging to different controls). Technically, this makes the graph of a system a multigraph.}. 
\end{dfn}

Since the nodes of the graph are labelled by the eigenstates $\ket{e_k}$, we use these labels also to name the edges that represent the transitions generated by the controls. The edge joining the vertices $\ket{e_a}$ and $\ket{e_b}$ is denoted by  $(a,b)$. For Hermitian controls, the transitions will necessarily happen in both directions, implying an undirected graph. This means, in particular, that both $(a,b)$ and $(b,a)$ refer to the same edge.

An important aspect of the graph are the energy gaps related to the graph edges $\Delta E_{a,b} := \left| E_{\ket{e_a}} - E_{\ket{e_b}} \right|$. If two transitions have the same energy gap $\Delta E_{a,b}$, they are called 'resonant'. Resonant transitions may pose a challenge to controllability since, depending on the local controls, it may be impossible to address them independently.

\begin{dfn}
Let $\mathcal{G}$ be the graph of a system. If two edges $(a,b), (c,d) \in E(\mathcal{G})$ belong to the same control $\hat{H}_j$ and have degenerate energy gaps $\Delta E_{a,b} = \Delta E_{c,d}$, then the two transitions represented by the edges are \textbf{coupled} to one another. Alternatively, if an edge $(a,b) \in E(\mathcal{G})$ is not coupled to any other transition belonging to its own control $\hat{H}_j$, the transition is said to be \textbf{decoupled}. 
\end{dfn}

This leads to the following theorem whose proof can be found in Ref. \cite{chambrion2009}:
\begin{theorem} \label{thm:graph_test}
Let $\mathcal{S}$ be a quantum system following \autoref{eqn:controlH} and let $\mathcal{G}$ be its associated graph. The system $\mathcal{S}$ is \textbf{controllable} if there exists a \textbf{connected} \textbf{subgraph} of $\mathcal{G}$ that contains\textbf{ all vertices} of $\mathcal{G}$ and \textbf{only decoupled transitions}. 
\end{theorem}

An undirected graph is said to be connected if for every two vertices there exists a chain of adjacent edges that creates a path between the two selected vertices. A subgraph of a graph $\mathcal{G}$ is a graph defined by a subset of the vertices $V(\mathcal{G})$ and a subset of the edges $E(\mathcal{G})$ that only link vertices in the subset. This means that a subgraph is a graph that we obtain by removing any number of the edges and vertices from the original graph.

The main benefit of the graph test in \autoref{thm:graph_test} is that it allows us to avoid calculating the full dynamical Lie algebra. The graph of a quantum system with Hilbert space dimension $n$ has $n$ vertices. 
To have a connected subgraph we need to find $n-1$ decoupled transitions such that they create a single connected component. This works quite efficiently if there are no resonant transitions in the system. 

In the following, we expand this graph test such that it also works with resonant transitions. To this end, we adapt graphical methods for controllability analysis in degenerate systems \cite{pozzoli2022} to systems with degenerate energy gaps driven by multiple controls. Since resonant transitions are very common in qubit arrays due to the multi-partite structure of the system, this modification opens the route to efficient analysis of controllability of qubit arrays.


\section{Graph test of controllability for coupled subsystems} \label{sec:algorithm}

In this section we outline the algorithm for a graph test for coupled subsystems with resonant transitions. In \autoref{ssec:notation-and-tools} we present methods to determine the graphical commutators already used in rotor systems \cite{leibscher2022} and we introduce the concept of subalgebras to treat resonant transitions driven by multiple controls. We show in \autoref{ssec:description-algorithms} that the graphical commutators can be calculated in a systematic way, avoiding the construction of the entire dynamical Lie algebra of a system. We illustrate the use of these methods in \autoref{ssec:simple-examples} with simple two-qubit examples.

\subsection{Resonant transitions and graphical commutators} \label{ssec:notation-and-tools}

We first present two concepts to decouple resonant transitions without calculating the complete Lie algebra of the system.
To this end, we introduce the generalized skew-Hermitian Pauli matrices
\begin{align} \label{eqn:pauli}
    \hat{G}_{j,k} &= e_{j,k} - e_{k,j}\,,  \nonumber\\
\hat{F}_{j,k} &= ie_{j,k} + ie_{k,j}\,, \\
\hat{D}_{j,k} &= ie_{j,j} - ie_{k,k}\,, \nonumber
\end{align}
where $e_{j,k}$ is the null matrix except for a 1 in the entry $(j,k)$. The commutators of these matrices with the skew-Hermitian drift are given by \cite{boscain2021classical, pozzoli2022}
\begin{align} \label{eqn:H0-commutators}
    [  i\hat{H}_0, \hat{G}_{j,k}] &= -\Delta E_{k,j} \hat{F}_{j,k}\,,  \nonumber\\
    [  i\hat{H}_0, \hat{F}_{j,k}] &= \Delta E_{k,j} \hat{G}_{j,k}\,.  
\end{align}
Consider a transition $(i,k)$ with transition matrix element 
\begin{equation} \label{eqn:coef1}
 \langle e_i | \hat{H}_j | e_k \rangle = \alpha - i \beta\,,
\end{equation}
where $\ket{e_i}$, $\ket{e_k}$ with $i\neq k$ are eigenstates of the drift Hamiltonian, $\hat{H}_0$, $\alpha, \beta \in \mathbb{R}$ and at least one of them is nonzero. 
If $(i,k)$ is a decoupled transition, the Lie algebra of the system contains the element
\begin{equation} \label{eqn:coef2}
   \hat{T}  =  \alpha \hat{F}_{i,k} + \beta \hat{G}_{i,k}\,,
\end{equation}
and the generalized skew-Hermitian Pauli matrices $\hat{G}_{i,k},\hat{F}_{i,k}$ and $\hat{D}_{i,k}$ are also elements of the Lie algebra $Lie(i \hat{H}_0,\hat{T})$, i.e., $dim[Lie(i \hat{H}_0,{\hat T})]=4$.

\begin{figure}[tbp]
\centering
\includegraphics[width=0.45\textwidth]{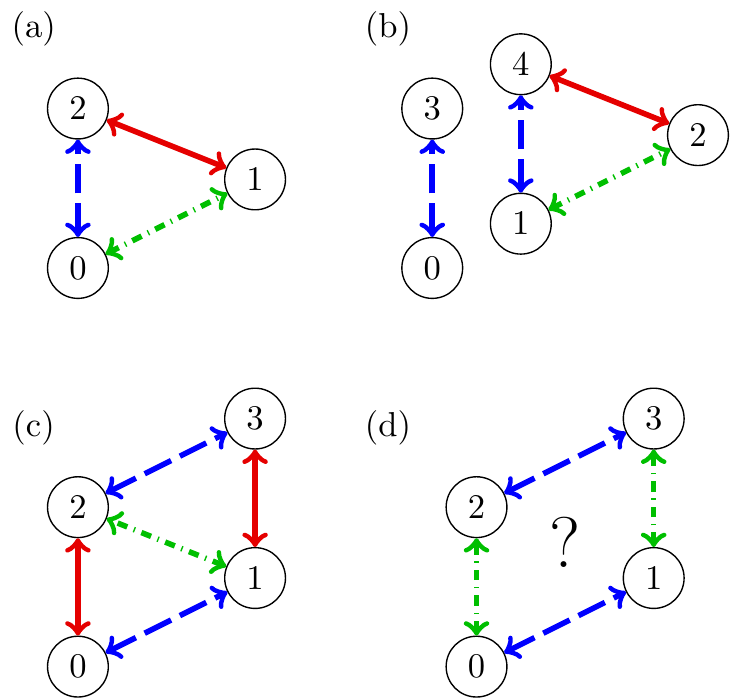}
 \caption{Examples for graphical commutators. The circles represent the vertices of the graph and the edges (transitions) are depicted by the arrows. Same-colour arrows represent coupled transitions. In all cases, the commutator between the transitions depicted by dashed blue and dash-dotted green arrows result in transitions depicted by solid red arrows. The four panels present the following cases: 
 \textbf{(a)} Graphical commutator between two uncoupled transitions $[(0,1),(0,2)] = (1,2)$
 \textbf{(b)} Graphical commutator between the coupled transitions $\{(0,3),(1,4)\}$ and the uncoupled transition $(1,2)$ results in the uncoupled transition $(2,4)$
 \textbf{(c)} Graphical commutator between the coupled transitions $\{(0,1),(2,3)\}$ and the uncoupled transition $(1,2)$ results in the coupled transition $\{(0,2), (1,3)\}$.
 \textbf{(d)} The commutator between the coupled transitions $\{(0,2),(1,3)\}$ and $\{(0,1),(2,3)\}$ cannot be determined graphically.}
 \label{fig:graph_comm}
\end{figure}

If a system contains the two decoupled transitions $\{(i,k), (i,l)\}$ with $\hat{T}_1=\alpha_1 \hat{F}_{i,k} + \beta_1 \hat{G}_{i,k}$ and $\hat{T}_2=\alpha_2 \hat{F}_{i,l} + \beta_2 \hat{G}_{i,l}$, respectively, and $i \neq l \neq k$, their commutator is of the form
$\hat{T}_3 := [\hat{T}_1 , \hat{T}_2] = \alpha_3 \hat{F}_{k,l} + \beta_3 \hat{G}_{k,l}$. The graph of the system thus has an additional edge $(k,l)$, as depicted in \autoref{fig:graph_comm}(a) for $i=0$, $k=1$ and $l=2$. For testing controllability, it is often sufficient to know that there exists a transition $(k,l)$, without calculating the coefficients $\alpha_3,\beta_3$. Since the existence of the transition $(k,l)$ can be deduced from the graph with edges $(i,k)$ and $(i,l)$, we refer to this operation as graphical commutator and denote it as $(k,l)=[(i,k),(i,l)]$.

Graphical commutators can be used as an efficient tool to decouple resonant transitions. This is shown in the example depicted in \autoref{fig:graph_comm}(b)
Consider transitions $\hat{T}_1 =  \alpha_1 \hat{G}_{1,2} + \beta_1 \hat{F}_{1,2}$ (dash-dotted green arrow in \autoref{fig:graph_comm}(b)) and $\hat{T}_2 = \alpha_2 \hat{F}_{0,3}+\beta_2 \hat{G}_{0,3} + \gamma_2 \hat{F}_{1,4} + \delta_2 \hat{G}_{1,4}$ (dashed blue arrows) with $\hat{T}_1$ decoupled (i.e. the transition $(1,2)$ is not coupled to any other one) and $\hat{T}_2$ consisting of the pair of coupled transitions $\{(0,3),(1,4)\}$, which are not coupled to any other transition. The graphical commutator between $\hat{T}_1$ and $\hat{T}_2$ has then only contributions corresponding to the transition $(2,4)$, depicted by the red arrow. The graph has thus the additional decoupled transition $(2,4)$. Furthermore, the graphical commutator between the decoupled transitions $(1,2)$ and $(2,4)$ is the decoupled transition $(1,4)$. In this example, the resonant transitions $\{(0,3), (1,4)\}$ can thus be decoupled by taking graphical commutators. 
 
In certain instances, the graphical commutators might not give enough information to decouple resonant transitions and determine the controllability of the system. Examples are shown in \autoref{fig:graph_comm}(c) and (d). In \autoref{fig:graph_comm}(c), the graphical commutator between the coupled transitions $\{(0,1),(2,3)\}$ (dashed blue arrows) and the decoupled transition $(1,2)$ (dash-dotted green arrows) results in the coupled transitions $\{(0,2),(1,3)\}$ (red arrows). This does not allow for decoupling any of the coupled transitions. In \autoref{fig:graph_comm}(d), two pairs of coupled transitions are considered, namely $\{(0,1),(2,3)\}$ (dashed blue arrows) and $\{(0,2),(1,3)\}$ (dash-dotted green arrows). The graphical commutators $[(0,1),(0,2)]$ and $[(1,3),(2,3)]$ both result in the transition $(1,2)$. Without further knowledge of the coefficients of the corresponding Lie algebra elements, it is not possible to determine whether the two terms result in a non-zero transition $(1,2)$, or whether they cancel each other. 

In such cases, we make use of an alternative procedure to decouple resonant transitions, which takes into account that in qubit systems, the same (coupled) transitions are often driven by different controls. Consider a system with drift
$\hat{H}_0$ and two controls that both drive the two resonant transitions $\{(1,2),(3,4)\}$. The Lie algebra of the system thus contains the terms
\begin{align}
    \hat{T}_4 &= \alpha_4 \hat{F}_{1,2}+\beta_4 \hat{G}_{1,2} +\gamma_4 \hat{G}_{3,4} + \delta_4 \hat{G}_{3,4}\,,
     \nonumber \\
    \hat{T}_5 &= \alpha_5 \hat{F}_{1,2}+\beta_5 \hat{G}_{1,2} +\gamma_5 \hat{G}_{3,4} + \delta_5 \hat{G}_{3,4}\,.
\end{align}
The two transitions $(1,2)$ and $(3,4)$ are decoupled if 
$\hat{T}_6 := \alpha_6 \hat{F}_{1,2}+\beta_6 \hat{G}_{1,2}$ is an element of the Lie algebra $Lie(  i\hat{H}_0, \hat{T}_4, \hat{T}_5)$ for any real $\alpha_6, \beta_6$ (at least one of them nonzero). This is true if and only if $dim\left[ Lie(  i\hat{H}_0, \hat{T}_4, \hat{T}_5)\right] = dim\left[ Lie(i \hat{H}_0, \hat{G}_{1,2}, \hat{G}_{3,4})\right] = \,7$, i.e., when the generated sub-algebra has maximum dimension (for a given number of transitions). 
Note that we have deliberately chosen transitions $\{(1,2), (3,4)\}$, which have no vertex in common, i.e., they are disjoint. Restricting ourselves to disjoint transitions, the maximum dimension of the generated sub-algebra is $3 n_t +1$, where $n_t$ is the number of transitions that are coupled. In order to determine, if a set of $n_t$ transitions driven by $n_t$ different controls is decoupled, it is thus sufficient to calculate the dimension of the sub-algebra with maximal dimension $3 n_t +1$. This is typically much smaller than the dimension of the Lie algebra of the complete system. If the transitions are not disjoint, the dimension of the relevant subalgebra scales quadratically with the number of transitions $n_t$. This, although feasible for a small number of transitions, would make the construction of the subalgebras more demanding

In the following we use both methods, i.e., graphical commutators and the calculation of the dimension of small sub-algebras, in order to decouple resonant transitions. This allows us to extend the graph test for controllability to coupled susbystems with resonant transitions.


\subsection{Algorithms for graph test of controllability for coupled subsystems with resonant transitions}\label{ssec:description-algorithms}

\begin{figure}[tbp]
\centering
\includegraphics[width=0.82\textwidth]{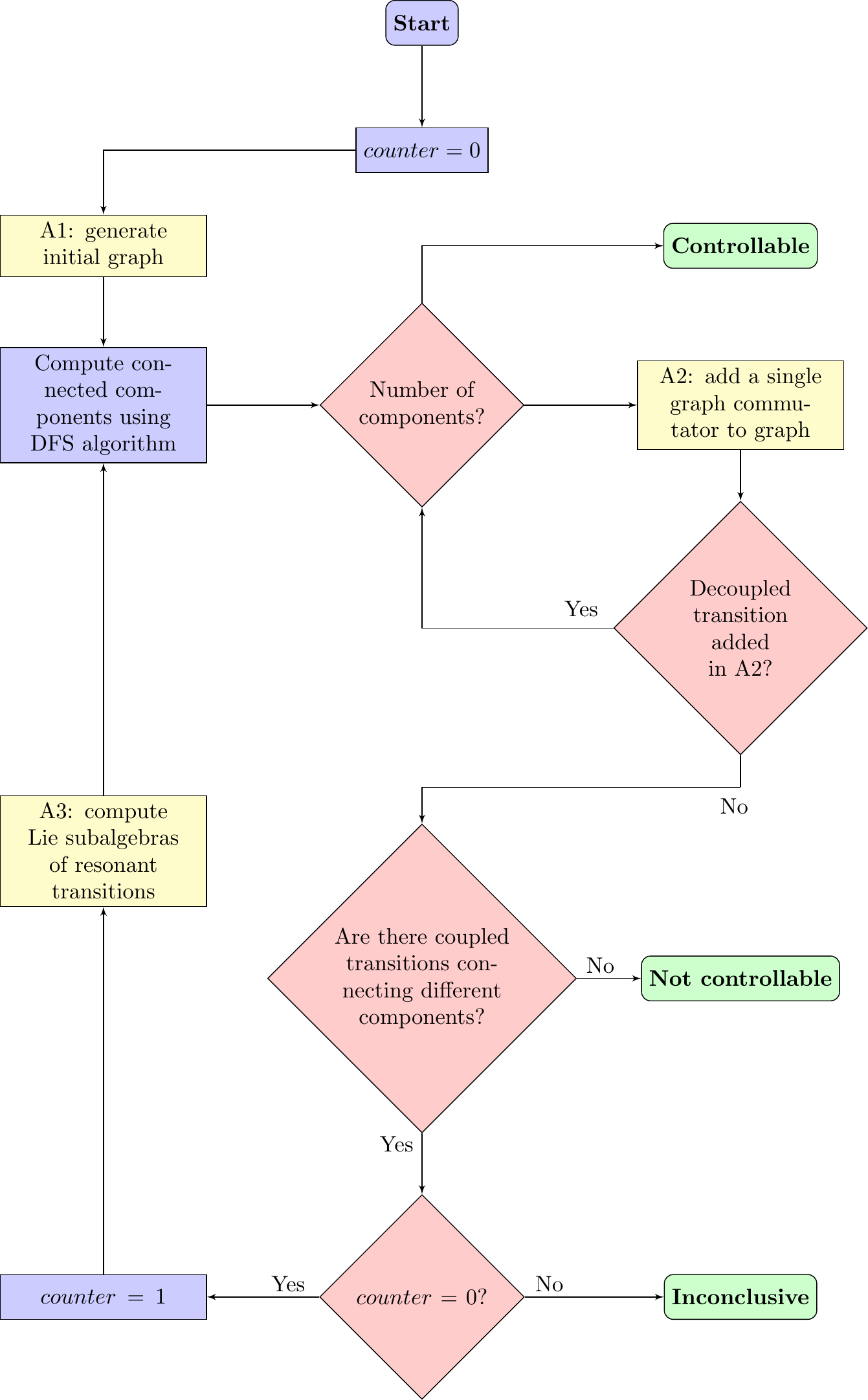}
 \caption{Flowchart representing the necessary steps and order in which the three main subroutines (A1-3) are used. The three possible results for this test are displayed in the three green round cells. }
 \label{fig:flowchart}
 
\end{figure}

Our graph test for controllability of a quantum system with resonant transitions is divided into several steps, depicted in \autoref{fig:flowchart}. The main output of the complete algorithm is a variable conveying whether the system is controllable, not controllable or whether the test remains inconclusive. 

\clearpage

\onecolumngrid   


\clearpage

\begin{algorithm}[tbp]
    
    Input: \newline
        $\bullet$ \textbf{\textit{H\_op}}: list including the drift $\hat{H}_0$ and all control operators $\hat{H}_j$ in the logical basis. 
        \newline
        $\bullet$ $\mathbf{\delta_H}$: minimum tolerance for a nonzero transition coefficient.
         \newline
        $\bullet$ $\mathbf{\delta_E}$: minimum tolerance for two energy gaps to be identical.
        
Diagonalize $\hat{H}_0$ and represent all control operators $\hat{H}_j$ in the eigenbasis of $\hat{H}_0$.   

Initialize \textit{decoupled-graph}, representing every eigenstante $\ket{e_a}$ of $\hat{H}_0$ as a vertex in the graph.   

\For{every control $\hat{H}_j$ in \textit{H\_op}}{
    \For{$0\leq a < b < \text{Hilbert space dimension}$}{
        \If{$\left|\bra{e_a}\hat{H}_j\ket{e_b}\right| > \delta_H$}
            {Add $(a,b)$ to \textit{transition-list.}
        }
    }  
    
    Sort \textit{transition-list} by the energy gaps $\Delta E_{a,b} := |E_a - E_b|$

    \For{every $(a,b)$ in \textit{transition-list}}{
        Add all transitions $(c,d)$ resonant to $(a,b)$ (i.e. $\left|\Delta E_{a,b}-\Delta E_{c,d}\right| < \delta_E$) to a new, separate list \textit{resonant-trans-ab} (including $(a,b)$).

        Add coefficients $\bra{e_c}\hat{H}_j\ket{e_d}$ of all resonant transitions $(c,d)$ into a new list \textit{resonant-coef-ab} (including $(a,b)$).

        \eIf{\textit{resonant-trans-ab} has more than one element}
        {
            Add \textit{resonant-trans-ab} as a new entry in the list \textit{coupled-transitions}.

            Add \textit{resonant-coef-ab} as a new entry in the list \textit{coupled-coefficients}.            
        }{
            Add \textit{resonant-trans-ab} as a decoupled transition in \textit{decoupled-graph}.    
        }    
        
        Remove all resonant transitions $(c,d)$ and $(a,b)$ from \textit{transition-list}.     
    }
}
   
    Output: \newline
        $\bullet$ \textbf{\textit{decoupled-graph}}: a dictionary containing all vertices of the graph (eigenstates of $\hat{H}_0$) and all decoupled transitions  as edges.
        \newline $\bullet$ \textbf{\textit{coupled-transitions}}: a list of lists containing all coupled transitions stored as dictionaries and sorted by their respective energy gap $\Delta E_{a,b}$. Resonant transitions from the same control with the same $\Delta E_{a,b}$ are stored in the same inner list since they are coupled. 
        \newline $\bullet$ \textbf{\textit{coupled-coefficients}}: a list of lists of the transition coefficients, matching the order of \textit{coupled-transitions}.

  \caption{Compute the initial graph of the system}
  \label{alg:A1}
\end{algorithm}

\begin{algorithm}[tbp]
Input: \newline
        $\bullet$ \textbf{\textit{decoupled-graph}}: from \autoref{alg:A1}.
        \newline $\bullet$ \textbf{\textit{graph-components}}: list containing the sets representing the connected components of \textit{decoupled-graph}. 
        \newline $\bullet$ \textbf{\textit{coupled-transitions}}: from \autoref{alg:A1}.

\textit{transition-added} = False

\textit{connecting-transitions-found} = False

$i=0$

\While{($i$ < length of \textit{coupled-transitions}) \textbf{and} (\textit{transition-added} = False)}{
    $i = i+1$

    \textit{element\_i} = \textit{coupled-transitions}$[i]$
    \newline {\color{gray} \# element\_i is a set of coupled transitions of the form $\{(a_0, a_1), (a_2, a_3), ...\}$}

    $j=0$

    \While{($j$ < length of \textit{element\_i}) \textbf{and} (\textit{transition-added} = False)}{
    
        $j=j+1$ 

        $(b_0, b_1) = $ \textit{element\_i}$[j]$

        \If{$b_0$ and $b_1$ belong to different components in \textit{graph-components}}{
    
            \textit{connecting-transitions-found} = True
    
            \textit{component\_0} = component in \textit{graph-components} containing $b_0$
    
            \textit{component\_1} = component in \textit{graph-components} containing $b_1$
    
            $k=0$
    
            \While{($k$ < number of transitions in \textit{decoupled-graph}) \textbf{and} (\textit{transition-added} = False)}{

                $k=k+1$
    
                $(c_0, c_1) = k\text{-th}$ transition in \textit{decoupled-graph}
    
                \If{($\{c_0, c_1\} \subset$ \textit{component\_0}) \textbf{or} ($\{c_0, c_1\} \subset$ \textit{component\_1})}{

                    \textit{graphical-commutator} = [\textit{element\_i}, $(b_0, b_1)$]
    
                    \If{\textit{graphical-commutator} is a single transition}{

                        \textit{transition-added} = True

                        Add \textit{graphical-commutator} to \textit{decoupled-graph}

                        Merge \textit{component\_0} and \textit{component\_1} in \textit{graph-components}
                    }
                }
            }
        }
    }
}

Output: 
    \newline $\bullet$ \textbf{\textit{decoupled-graph}}: updated after the routine.
    \newline $\bullet$ \textbf{\textit{graph-components}}: updated after the routine.
    \newline $\bullet$ \textbf{\textit{transition-added}}: True if a transition was added during this subroutine to \textit{decoupled-graph}; False otherwise.
    \newline $\bullet$ \textbf{\textit{connecting-transitions-found}}: True if at least one transition in \textit{coupled-transitions} connects different components; False otherwise.
    
  \caption{Add a single decoupled transition using graphical commutators}
  \label{alg:A2}
\end{algorithm}

\begin{algorithm}[tbp] 
    Input: 
    \newline $\bullet$ \textbf{\textit{decoupled-graph}}: from \autoref{alg:A2}.
    \newline $\bullet$ \textbf{\textit{graph-components}}: from \autoref{alg:A2}. 
    \newline $\bullet$ \textbf{\textit{coupled-transitions}}: from \autoref{alg:A1}.
    \newline $\bullet$ \textbf{\textit{coupled-coefficients}}: from \autoref{alg:A1}.

\For{every set of resonant transitions $\{(a_0, a_1), (a_2, a_3), ...\}$ in \textit{coupled-transitions}}{
    \textit{resonant-set} = $\{(a_0, a_1), (a_2, a_3), ...\}$
    
    \If{\textit{resonant-set} contains only disjoint transitions}{
        \If{\textit{resonant-set} connects different components in \textit{graph-components}}{
            \If{\textit{resonant-set} appears multiple times in \textit{coupled-transitions}}{
                Set $m$ as the number of times \textit{resonant-set} is generated by different controls.
                
                Set $n_t$ as the number of coupled transitions in \textit{resonant-set} 
                
                \For{$0\leq j < m$}{

                    Using the coefficients in \textit{coupled-coefficients} related to the $j$-th instance of \textit{resonant-set}, generate an array \textit{T-array-j} to represent the transition.
                    \newline {\color{gray} \#  $\hat{T}_j = \alpha_{a_0, a_1}^{(j)} \hat{F}_{a_0, a_1} + \beta_{a_0, a_1}^{(j)} \hat{G}_{a_0, a_1} + \alpha_{a_2, a_3}^{(j)} \hat{F}_{a_2, a_3} + \beta_{a_2, a_3}^{(j)} \hat{G}_{a_2, a_3} + ...$}
                    \newline {\color{gray} \# \textit{T-array-j} $= \left( \alpha_{a_0, a_1}^{(j)} , \beta_{a_0, a_1}^{(j)}, 0,  \alpha_{a_2, a_3}^{(j)}, \beta_{a_2, a_3}^{(j)}, 0,  ... \right)$}                         
                 }                   
                \If{$dim \left(Lie\left\{i\hat{H}_0, \textit{T-array-0}, ..., \textit{T-array-m}\right\}\right) = 3 n_t +1 $}{
                    Add every transition in \textit{resonant-set} to \textit{decoupled-graph}
                }    
            }
        }
    }
}

Output: 
    \newline $\bullet$ \textbf{\textit{decoupled-graph}}: updated after the routine.

  \caption{Subroutine to compute subalgebras of repeated resonant transitions}
  \label{alg:A3}
\end{algorithm}

\twocolumngrid

\clearpage

As can be seen in \autoref{fig:flowchart}, the first stage is creating the initial graph of the quantum system (A1).  
The step-by-step definition is found in \autoref{alg:A1}. To compute this graph, we diagonalize the drift $\hat{H}_0$, taking its eigenstates as the vertices of the graph. Then, we determine the nonzero transitions $\bra{e_a}\hat{H}_j\ket{e_b}$ for all $m$ controls and take these to be the edges $(a,b)$ of the graph. Numerically, we take into account all transitions that are larger or equal than a certain tolerance, $\delta_H > 0$, i.e.,
\begin{equation}
    \left|\bra{e_a}\hat{H}_j\ket{e_b}\right| \geq \delta_H \quad \longrightarrow \quad (a,b)\,.
\end{equation}

The next step is to determine which transitions are coupled. As defined in \autoref{sec:controllability}, two or more transitions are coupled if and only if they have the same energy gap and are generated by the same control. Any transition not coupled to any other one is by definition a decoupled transition. Note that two transitions generated by different controls can never be coupled to one another. 
To numerically compare the energy gaps, we define a minimal tolerance $\delta_E > 0$ such that two transitions driven by the same control, $(a,b)$ and $(c,d)$, are considered coupled if $\left| \Delta E_{a,b} - \Delta E_{c,d} \right| \leq \delta_E$. 
\noindent We sort all the transitions of each control according to the energy gaps $\Delta E_{a,b}$ and separate them depending on whether they are coupled or not. 
If a coupled transition $(a,b)^{(j)}$ is generated by multiple controls $\hat{H}_j$, we also calculate and store the corresponding transition coefficients $\bra{e_a}\hat{H}_j\ket{e_b}=\alpha_{a,b}^{(j)} + \beta_{a,b}^{(j)} i$. Note that here, we have introduced the notation $(a,b)^{(j)}$ to indicate that the transition $(a,b)$ is driven by the control $\hat {H}_j$.

 The output of \autoref{alg:A1} encompasses the \textit{decoupled-graph}, the \textit{coupled-transitions} and the \textit{coupled-coefficients}. 
The first output, \textit{decoupled-graph}, is a dictionary that contains all vertices (eigenstates) and all decoupled edges (decoupled transitions) of the systems graph. These are the elements that we can use directly to test the controllability of the system using \autoref{thm:graph_test}. In the second output, \textit{coupled-transitions}, every set of coupled transitions is stored as a list of tuples $(a,b)$. Coupled transitions cannot be immediately used for the test in \autoref{thm:graph_test}, but are necessary for generating additional decoupled transitions, by using graphical commutators or subalgebras of resonant transitions. 
The last output variable \textit{coupled-coefficients} consists of the entries $\bra{e_a}\hat{H}_j\ket{e_b}=\alpha_{a,b}^{(j)} + \beta_{a,b}^{(j)} i$ for every coupled transition $(a,b)^{(j)}$ which are required for calculating the dimension of the subalgebras. 

Once the initial graph of the system has been computed, we check if the graph containing only decoupled transitions (\textit{decoupled-graph}) is connected.  To do so, we determine the number of connected components in the graph. In terms of graph theory, a component is a connected subgraph that it is not contained in any larger connected subgraph. In other words, each of the connected parts into which we can divide a graph is called a component of the graph. Therefore, the graph is connected if and only if it has exactly one component. One of the many possible ways to count the number of components is to use the depth-first search (DFS) algorithm\footnote{The DFS algorithm is a common recursive algorithm to explore explore systematically all vertices in a graph. It uses an exhaustive search by traversing down a chain of edges, or by backtracking when not possible.}. If the \textit{decoupled-graph} is already connected, we can stop the routine and state that the system is indeed controllable. 

If the graph is not connected, then \autoref{alg:A2} is called. The aim of this algorithm is to search for an additional edge of the graph by using graphical commutators. As explained in \autoref{sec:controllability}, the graphical commutators of a coupled transition and a decoupled transition may generate a new decoupled transition.
 Note that, instead of identifying and adding every possible combination of commutators between a decoupled transition and a set of coupled transitions, we are only interested in  commutators that result in a decoupled transition which connects different components of the graph and therefore potentially creates a connected graph. That is, in \autoref{alg:A2} we compute the commutator of a coupled transition that connects two separate components of \textit{decoupled-graph} with a decoupled transition in one of the two components.
  If the result of this commutator is a decoupled transition, then this new edge will connect the two components in \textit{decoupled-graph}, merging them into a single component.  If a new decoupled transition is added to the \textit{decoupled-graph} during this step, the number of connected components is counted again to determine if the graph is now connected. 

If the number of components is still larger than one, i.e. if the \textit{decoupled-graph} is not connected after \autoref{alg:A2}, and no new transition was added during its last call, then we check if there is any transition in \textit{coupled-transitions} that connects different components of \textit{decoupled-graph}. If this is not the case, the routine is stopped. If all transitions, coupled and decoupled, only connect vertices within the same components, then all possible commutators result in transitions within the same components. The number of components will stay constant and the graph will never be connected. This implies that the system is not controllable. 

However, if there exists at least one coupled transition connecting different components, then this argument is no longer valid and the controllability of the system is not yet decided. 
For these cases, we use the \textit{coupled-coefficients} that are related to the \textit{coupled-transitions} to obtain a more conclusive answer.

This step is carried out by \autoref{alg:A3}. It builds upon the updated \textit{decoupled-graph} from \autoref{alg:A2} and tries to add new decoupled transitions by computing the low-dimensional subalgebras of resonant transitions driven by different controls. 
First, the algorithm selects the transitions from \textit{coupled-transitions} that connect different components. 
In particular, we are exclusively interested in sets of disjoint transitions, such that the dimension of the associated subalgebras remains low. Next, the algorithm finds out if those transitions have a multiplicity equal to or higher than 2, that is, if the same sets of coupled transitions $\left\{(a_0,a_1), (a_2,a_3), ...\right\}^{(j)}$ are generated multiple times by different controls $\hat{H}_j$. Following the reasoning presented in \autoref{ssec:notation-and-tools}, we try to generate the full associated subalgebra of the \textit{coupled-transitions}, $\left\{(a_0,a_1), (a_2,a_3), ...\right\}$, by using their respective \textit{coupled-coefficients}. If a subalgebra with maximal dimension is generated, we take all transitions $\left\{(a_0,a_1), (a_2,a_3), ...\right\}$ as effectively decoupled. This allows us to add every transition $(a_k, a_l)$ as a decoupled edge to \textit{decoupled-graph}. Note that in \autoref{alg:A3}, we compute the Lie subalgebras of the resonant transitions including $i\hat{H}_0$, where we deduce the commutators $\left[ i\hat{H}_0, \left\{(0,1), (2,3)\right\}^{(j)} \right]$ using \autoref{eqn:H0-commutators}. This makes these calculations simple and encompassed within a vector space of dimension $3n_t$, with $n_t$ being the number of transitions in $\left\{(a_0,a_1), (a_2,a_3), ...\right\}$. 

Finally, we compute the new number of components in \textit{decoupled-graph} using DFS once again. If the graph is still not connected, we use \autoref{alg:A2} to add as many graphical commutators as possible to connect the remaining components. If the graph ends up with one single component, the system is controllable. If there are more components and there are no transitions in \textit{coupled-transitions} that connect any of them, then the system is proven to be not controllable. If none of these two hold to be true, then the controllability test remains inconclusive. Indeed, unlike the dynamical Lie algebra method, the test might yield no definitive answer. If a graph is not connected, we cannot ensure that the system is not controllable, generally speaking. It might be that the graph can be connected after computing a large number of commutators or that it will remain forever not connected. Calculating this quantity of commutators is tantamount to computing the dynamical Lie algebra and would therefore be unfeasible for relatively large systems.


\subsection{Illustrative examples}\label{ssec:simple-examples}

Here we present two examples constructed to showcase how the algorithms described in \autoref{ssec:description-algorithms} work. 
 \begin{figure}[tbp]
\centering
\includegraphics[width=0.45\textwidth]{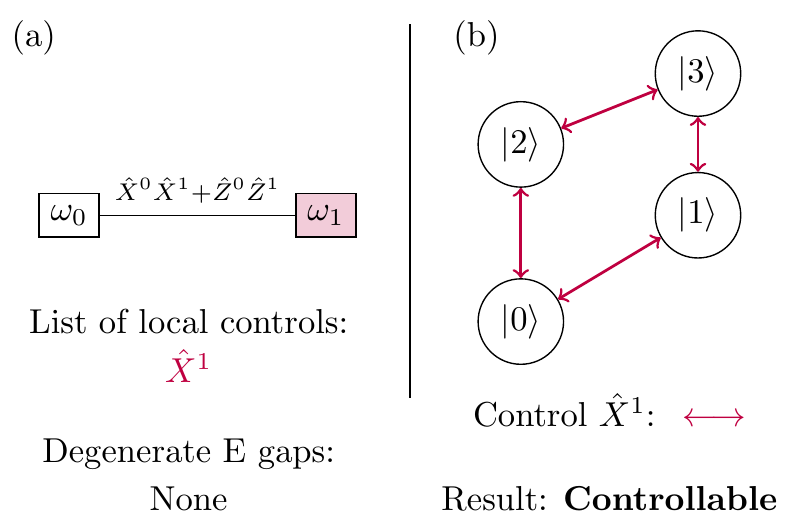}
 \caption{\textbf{(a)} Two-qubit system with Hamiltonian (\ref{eqn:2qubit_XXZZ}), where $\omega_0 = 5$ GHz, $\omega_1 = 5.5$ GHz, $J = 150$ MHz. \textbf{(b)} Graph of the system. All transitions are decoupled, making the system controllable by Theorem \ref{thm:graph_test}. }
 \label{fig:2qubit_XXZZ}
\end{figure}
We start by considering the two-qubit system shown in 
\autoref{fig:2qubit_XXZZ}(a), described by the Hamiltonian 
 \begin{equation}\label{eqn:2qubit_XXZZ}
    \hat{H}_{2A}(t) = \hat{H}_0 + u_1(t) \hat{H}_1
\end{equation}    
with the drift
\begin{equation}\label{eqn:2qubit_XXZZ_drift}
  \hat{H}_0 = \sum_{i=0}^1 -\frac{\omega_i}{2} \hat{Z}^i + J(\hat{X}^0\hat{X}^1 +\hat{Z}^0 \hat{Z}^1)\,,
 \end{equation}
and a single local control $\hat{H}_1 = \hat{X}^1$ acting on qubit 1.
Here, $\omega_i$ are the natural frequencies of the qubits and $J$ is the coupling strength. For simplicity, we use the notation $\hat{X}^j = \hat{\sigma}_x^j$ for the Pauli matrices. Algorithm \ref{alg:A1} results in the graph shown in \autoref{fig:2qubit_XXZZ}(b) with the circular vertices representing the eigenstates of $\hat{H}_0$. The local control $\hat{H}_1$ gives rise to four non-zero transitions, which form the edges of the graph, denoted by the arrows in \autoref{fig:2qubit_XXZZ}(b). The different lengths of the arrows indicate that all four transitions have different energy gaps and are thus decoupled. The actual controllability test for this example consists simply of \autoref{alg:A1} and the DFS algorithm, which proves that the graph is connected and the system is controllable. In other words, from every vertex $i=0, ...3$, every other vertex of the graph can be reached by following the edges belonging to decoupled transition, as seen in \autoref{fig:2qubit_XXZZ}(b). 

 \begin{figure}[tbp]
\centering
\includegraphics[width=0.45\textwidth]{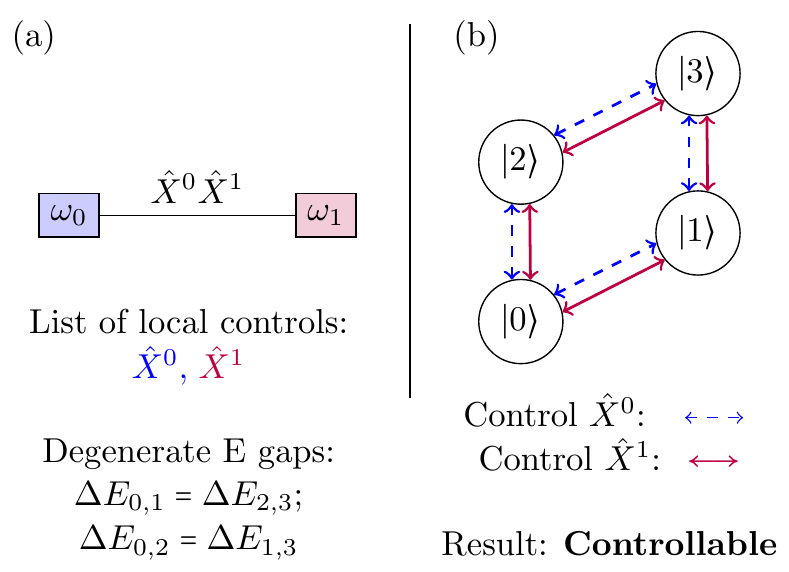}
 \caption{\textbf{(a)} Two-qubit system with Hamiltonian (\ref{eqn:2qubit_XX}) where $\omega_0 = 5$GHz, $\omega_1 = 5.5$ GHz, $J = 150$ MHz. Both local controls are necessary for the system to be controllable. \textbf{(b)} Graph of the system. Theorem \ref{thm:graph_test} is not directly applicable due to the coupled transitions. A study of the coefficients reveals that it is possible to decouple all transitions.}
 \label{fig:2qubit_XX}
\end{figure}

A second example is illustrated in \autoref{fig:2qubit_XX}(a).
Here, the Hamiltonian is given by 
\begin{equation}\label{eqn:2qubit_XX}
    \hat{H}_{2B}(t) = \hat{H}_0 + \sum_{j=1}^2 u_j(t) \hat{H}_j
\end{equation}
with the drift
\begin{equation}
    \hat{H}_0  = \sum_{i=0}^1 -\frac{\omega_i}{2} \hat{Z}^i + J\hat{X}^0\hat{X}^1\,,
\end{equation}
and two local controls
\begin{equation}
    \hat{H}_1 = \hat{X}^1 \quad \quad \mbox{and} \quad \quad 
    \hat{H}_2 = \hat{X}^0
\end{equation}
acting on the qubits 1 and 0, respectively. The main difference between \autoref{eqn:2qubit_XX} and \autoref{eqn:2qubit_XXZZ} is the omission of $\hat{Z}^0\hat{Z}^1$ in the drift Hamiltonian. This severely impacts the graph, as seen in \autoref{fig:2qubit_XX}. The symmetries that had previously been broken by including $\hat{Z}^0\hat{Z}^1$ appear in the energy gap degeneracies, $\Delta E_{0,1} = \Delta E_{2,3}$ and $\Delta E_{0,2} = \Delta E_{1,3}$. 
There are no decoupled transitions and the number of 
components in the initial graph is equal to four. 
A second difference is the second local control. The transitions driven by the controls $\hat{X}^1$ (which is the same local control as in \autoref{eqn:2qubit_XXZZ}) and $\hat{X}^0$ consist of two pairs of coupled transitions $\{(0,1), (2,3)\}$ and $\{(0,2), (1,3)\}$ shown by purple solid and dashed blue arrows respectively in \autoref{fig:2qubit_XX}(b).

\begin{table}[t]
\centering
\begin{tabular}{l|cc|cc}
\toprule
& \multicolumn{4}{c}{Transition parameters} \\
\cmidrule(r){2-5}
 & $(0,1)$ & $(2,3)$  & $(0,2)$ & $(1,3)$\\
\midrule \small
$\hat{X}^0$ & $0.967$ & $0.967$  & $-0.253$ & $0.253$\\
$\hat{X}^1$ & $-0.281$ & $0.281$  & $-0.960$ & $-0.960$\\
\bottomrule
\end{tabular}
\caption{Control coefficients $\bra{e_a} \hat{H}_j \ket{e_b}$ for the system (\ref{eqn:2qubit_XX}). The coupled transitions are $\{(0,1), (2,3)\}$ and $\{(0,2), (1,3)\}$. They appear with double multiplicity, once for each control. 
}
\label{tab:coef_2qubits_XX}
\end{table}

In this example, graphical commutators do not yield any new transition that we could add to the graph. In fact, we can only take commutators of pairs of coupled transitions, analogously to the case depicted in \autoref{fig:graph_comm}(d). Since there is no definitive answer, graphical commutators are not useful in this particular instance. However, since there exist coupled transitions which connect different components of the graph, \autoref{alg:A3} can be applied. The coefficients $\bra{e_a} \hat{H}_j \ket{e_b} =\alpha_{a,b}^{(j)} + \beta_{a,b}^{(j)} i  $ of the coupled transitions are shown in \autoref{tab:coef_2qubits_XX}. Note that in this example, the coefficients are real,  $\beta_{a,b}^{(j)}=0$. The Lie algebra thus contains the terms
\begin{eqnarray}
    \hat{T_0}=\alpha_{0,1}^{(0)} {\hat F}_{0,1} + \alpha_{2,3}^{(0)} {\hat F}_{2,3}\,, \nonumber \\
    \hat{T_1}= \alpha_{0,1}^{(1)} {\hat F}_{0,1} + \alpha_{2,3}^{(1)} {\hat F}_{2,3}\,.
\end{eqnarray}
It is immediate to see that we can isolate the term $\hat{F}_{2,3}$ by a linear combination of ${\hat T}_0$ and ${\hat T}_1$. This means that all operators $\hat{F}_{2,3}, \hat{G}_{2,3}, \hat{D}_{2,3}$ are elements of the subalgebra. Similarly, the opposite linear combination can isolate the element $\hat{F}_{0,1}$, adding  $\hat{F}_{0,1}, \hat{G}_{0,1}, \hat{D}_{0,1}$ to the generated subalgebra. This means that the dimension of $Lie\left(i\hat{H}_0,\hat{T}_0,\hat{T}_1\right) = 7$  (including the contribution of $i\hat{H}_0$) is maximal and thus the transitions $(0,1)$ and $(2,3)$ can be decoupled. Analogously, $(0,2)$ and $(1,3)$ may be decoupled as well. This means that all the edges are decoupled in the graph of \autoref{fig:2qubit_XX}. By virtue of \autoref{thm:graph_test}, this implies controllability of the system.


\section{Results}\label{sec:results}

In this section, we examine qubit systems which are used for quantum computing. With the algorithm presented in \autoref{sec:algorithm}, we can efficiently prove if a given qubit array is controllable. We present three different examples based on the IBM five-qubit array \textit{ibmq\_quito} \cite{IBMquito}. We do not use the exact parameters of this system, but simply its configuration and realistic parameters (except for the natural frequencies which correspond to the real data \cite{IBMquito}). 

The Hamiltonian of an array similar in structure to \textit{ibmq\_quito} can be expressed as:
\begin{equation}\label{eqn:quitoH}
    \hat{H}^{\textit{quito}}(t) = -\sum_{j=0}^4 \frac{\omega_j}{2} \hat{Z}^j + \hat{H}_c^{\textit{quito}} +\sum_{k=1}^m u_k (t) \hat{H}_k\,,
\end{equation}
where $u_k (t)$ represent local controls and the couplings
$\hat{H}_c^{\textit{quito}}$ are of the form
\begin{equation}\label{eqn:quitoHc}
    \hat{H}_c^{\textit{quito}} =  J_{0,1}\hat{H}_{0,1} + J_{1,2}\hat{H}_{1,2} + J_{1,3}\hat{H}_{1,3} + J_{3,4}\hat{H}_{3,4} 
\end{equation}
with each $\hat{H}_{j,k}$ representing an entangling (time-independent) coupling between the qubits $j$ and $k$. The following three examples are based on \autoref{eqn:quitoH} for different types of couplings and local controls. We will prove that while the original IBM design is indeed controllable there exist options that require fewer resources for controllability.

The set of parameters, including the natural frequencies $\omega_j$ and the coupling strengths $J_{i,j}$ are found in \autoref{tbl:quito_coef}. We use these parameters for all examples presented in this section.
\begin{table}[tb]
\centering
\begin{tabular}{ccccc}
\toprule
& \multicolumn{4}{c}{Coupling strengths (MHz)} \\
\cmidrule(r){2-5}
 & $J_{0,1}$ & $J_{1,2}$ & $J_{1,3}$ & $J_{3,4}$ \\
 & $100$ & $250$ & $170$ & $300$\\
\cmidrule(r){2-5}
\multicolumn{5}{c}{Qubit frequencies (GHz)} \\
\midrule 
$\omega_0$ & $\omega_1$ & $\omega_2$ & $\omega_3$ & $\omega_4$ \\
$5.301$ & $5.081$ & $5.322$  & $5.164$ & $5.052$\\
\bottomrule
\end{tabular}
\caption{Parameters for the system (\ref{eqn:quitoH}). The qubit frequencies are those of \cite{IBMquito}. The coupling strengths have been chosen to be in the range of hundreds of MHz, common for this type of qubits.}
\label{tbl:quito_coef}
\end{table}


\subsection{Example A: five-qubit system similar to IBM's \textit{ibmq\_quito}} \label{ssec: exampleA}

\begin{figure}[tbp]
\centering
\includegraphics[width=0.4\textwidth]{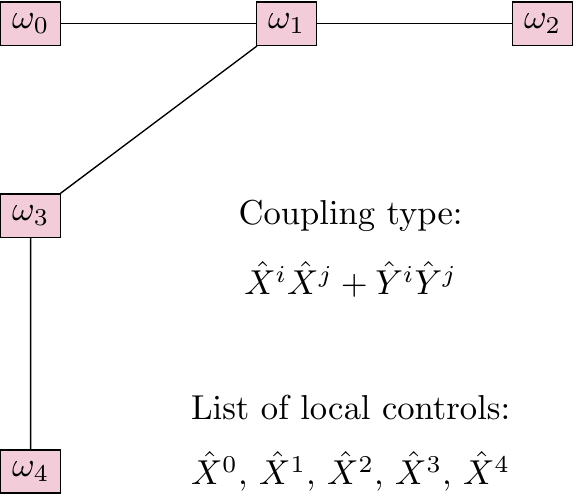}
 \caption{Five-qubit system inspired by IBM's \textit{ibmq\_quito} with Hamiltonian found in \autoref{eqn:quitoH}. 
 The rectangles present the qubits with natural frequencies $\omega_j$, $j=1-5$, and the lines indicate the couplings between the qubits. 
 Local controls are to be assumed in every qubit in the form of $\hat{X}^j$.}
 \label{fig:quito_original}
\end{figure}

This example mimics the qubit arrangement of \textit{ibmq\_quito}. The static couplings in \autoref{eqn:quitoHc} are \cite{roth2017analysis}
\begin{equation}
    \hat{H}_{ij} = J_{i,j} \left(\hat{X}^i\hat{X}^j+\hat{Y}^i\hat{Y}^j\right)\,.
\end{equation}
A local control $\hat{H}_j=\hat{X}^j$ is added to every qubit. The system diagram is shown in \autoref{fig:quito_original}. 

For a five-qubit system, there is a total of 32 vertices. After running \autoref{alg:A2} for this example with an energy gap tolerance of $\delta_E = 0.01$GHz, we obtain an initial graph with only 7 decoupled edges, resulting in a total of 25 different connected components. Use of \autoref{alg:A2} alone is not enough in this case, since the graphical commutators do not yield any new results. Executing \autoref{alg:A3}, a total of 192 new edges are added to the graph. These turn out to be more than sufficient to connect all the components and achieve a connected graph. 

We can thus conclude that the system presented in \autoref{fig:quito_original} is controllable and therefore suitable to perform any unitary operation. In the following examples we investigate whether the five-qubit array is also controllable if the number of local controls is reduced. 


\subsection{Example B: five-qubit system with reduced number of controls} \label{ssec: exampleB}

 \begin{figure}[tbp]
\centering
\includegraphics[width=0.4\textwidth]{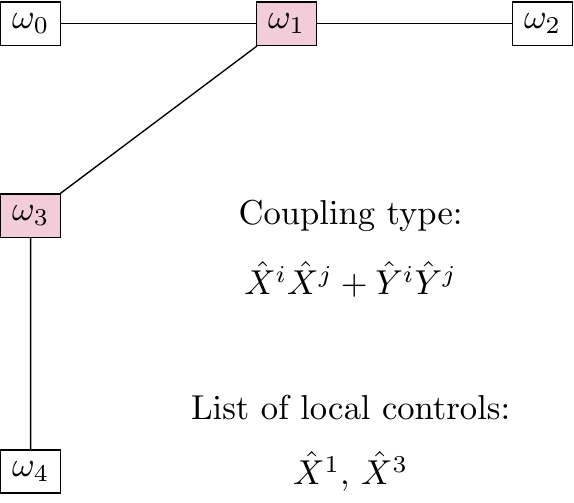}
 \caption{Same as \autoref{fig:quito_original} with fewer controls, located on qubits 1 and 3 (pink rectangles). }
 \label{fig:quito_fewer}
\end{figure}

Next, we investigate the controllability of a five-qubit system with a reduced number of local controls. For the drift Hamiltonian, we consider the same five-qubit system as before with all parameters as listed in \autoref{tbl:quito_coef} but decrease the number of local controls to two. This is depicted in \autoref{fig:quito_fewer}, which represents one of the ten different scenarios with two local controls $\hat{H}_j=\hat{X}^j$. 

Running \autoref{alg:A1} we obtain identical results to those in \autoref{ssec: exampleA}: a graph with only 7 decoupled transitions and 25 different connected components. Again, it is crucial to use \autoref{alg:A3} to make the graph connected. It may come as a surprise that the number of new edges that we can add is again 192. In other words, the number of new decoupled edges has not been reduced by removing three of the local controls. With the new decoupled edges included, the system is indeed controllable. 

We could attempt to take this study further to single local control. However, with a single local control $\hat{X}^j$, we always obtain an inconclusive result, no matter on which qubit we apply it. Although this hints to the possibility of the system not being controllable, the algorithm cannot conclusively determine it, similarly to the case shown in \autoref{fig:2qubit_XX}. 
In that two-qubit example we can prove using the Lie algebra method that a single local control $\hat{X}^j$ is never sufficient. We therefore conjecture that the least number of local controls required for the system to be controllable is two.

Since two local controls is the minimum number to obtain a controllable system, we now study how the position of the control affects controllability. We categorize all possible arrangements of two local controls $\hat{X}^j$ by their distance.  For example, the local controls for qubits 1 and 3, shown in \autoref{fig:quito_fewer}, have distance equal to 1 since they are direct neighbors connected by a single coupling. Analogously, we can set local controls separated by two couplings (e.g. on qubit 2 and 3) that have distance 2, or local controls for qubits separated by three couplings (e.g qubits 0 and 4) that have distance 3.
The results of the controllability tests for all these possible arrangements of two local controls are shown in \autoref{tbl:quito_2control}.  

\begin{table}[tbp]
\centering
\begin{tabular}{ccc}
\toprule
Distance & Controllability &  $\#$ cases \\
\midrule 
1& Controllable & 4\\
2& Inconclusive & 4\\
3& Controllable & 2\\
\bottomrule
\end{tabular}
\caption{Study of controllability for a five-qubit system similar to \textit{ibmq\_quito} but with only two local controls. }
\label{tbl:quito_2control}
\end{table}

Accordingly, the five-qubit system is controllable with two local controls if they have either distance one or three, independent of the position of the controls themselves. 
On the other hand, if the controls are at distance 2, the algorithm returns an inconclusive answer. This implies that \autoref{alg:A3} cannot decouple enough transitions for the graph to be connected. In fact, it turns out that no new transitions could be added at all. 
This is because the coefficients $\bra{e_a}\hat{H}_j\ket{e_b}$ of the different transitions depend on the position of the local control that drives them. By shifting a local control to a neighboring qubit, it is common to see a change in some of the coefficients. The transition coefficients of same-type controls for a set of coupled transitions tend to be linearly independent when the controls are shifted by an odd number of qubits and dependent in the even case. This allows us to generate maximal subalgebras only for controls that are at an odd qubit-distance, turning the coupled transitions into decoupled transitions that can eventually make the graph connected.


\subsection{Example C: five-qubit system with single local control}\label{ss:minimal-quito}

 \begin{figure}[tbp]
\centering
\includegraphics[width=0.4\textwidth]{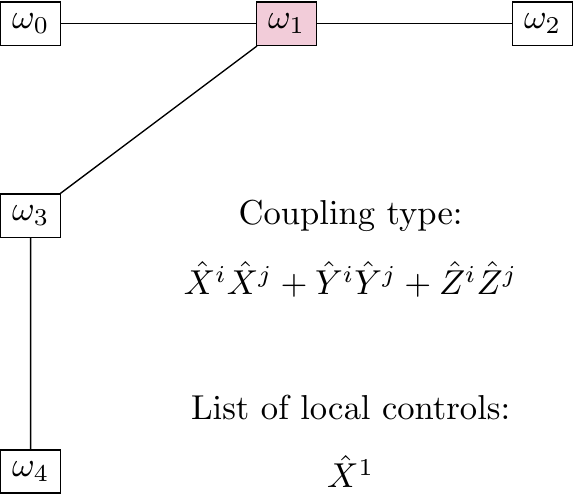}
 \caption{Diagram introducing a modified version of IBM's \textit{ibmq\_quito} with $\hat{Z}\hat{Z}$ contributions added to every coupling and all local controls removed except for $\hat{X}^1$.}
 \label{fig:quito_minimal}
\end{figure}

To conclude the list of examples, we demonstrate that the number of local controls that is required for the system to be controllable also depends on the type for couplings connecting the qubits. We showcase a system that is controllable with a minimal number of local controls, i.e., a single local control. The main difference between this example (shown in \autoref{fig:quito_minimal}) and the previous ones (displayed in \autoref{fig:quito_original} and \ref{fig:quito_fewer}) is that the couplings 
are of the form
\begin{equation}
   \hat{H}_{ij} = J_{i,j} \left(\hat{X}^i\hat{X}^j+\hat{Y}^i\hat{Y}^j +\hat{Z}^i\hat{Z}^j \right)\,,
\end{equation}
i.e., they have an extra term $\hat{Z}^i\hat{Z}^j$ compared to the couplings considered in the examples before. These kinds of couplings can be achieved in systems of superconducting qubits by harnessing both capacitive and inductive interactions between the qubits. Executing \autoref{alg:A1} with $\delta_E = 0.1$GHz, we obtain a total of 24 initially decoupled edges and a total of 14 connected components. This is already more promising than the examples in \autoref{ssec: exampleA} and \autoref{ssec: exampleB}. Furthermore, use of graphical commutators results in the system being controllable, ending up with 37 decoupled edges that are sufficient to make the graph connected. The controllability analysis thus shows that with the $\hat{Z}^i\hat{Z}^j$ addition, the system is controllable with only one local control. In \autoref{fig:quito_minimal}, we have added a local control on qubit 1, but additional tests show that the same statement is true for any single local control $\hat{X}^j$, independent of its position at qubit 0 to 4. Therefore, this controllable system has a minimal number of local controls at the cost of using a more complex type of two-qubit couplings.

The effect of couplings of the form $\hat{Z}^i\hat{Z}^j$ on the graph of the system can be understood by considering the simple two-qubit systems shown in  \autoref{fig:2qubit_XXZZ} and \autoref{fig:2qubit_XX}. In \autoref{fig:2qubit_XX}, couplings of the form $\hat{X}^i \hat{X}^j$ result in two pairs of resonant (coupled) transitions for each local control. Therefore, two local controls are required to decouple the transitions and ensure controllability. In \autoref{fig:2qubit_XXZZ}, a single control $\hat{X}$ is sufficient due to the $\hat{Z}^0\hat{Z}^1$ term in its coupling. This term is already diagonal in the eigenbasis of the free qubits, which means that it does not mix the eigenstates of the free qubits. However, it shifts their eigenenergies, such that all four vertices of the graph correspond to different transition frequencies and thus a single local control corresponds to four decoupled transitions, making the system controllable. A similar effect happens in the present five-qubit example, where the couplings have an off-diagonal contribution $\hat{X}^i\hat{X}^j+\hat{Y}^i\hat{Y}^j$ and a diagonal part $\hat{Z}^i\hat{Z}^j$. The former mixes the free-qubit eigenstates and in turn creates more edges in the graph with the inclusion of a local control like $\hat{X}^j$. The latter changes some resonant energy gaps and decouples the transitions associated to them. The combination of these two effects achieves controllability with a minimal number of local controls, as seen in \autoref{fig:quito_minimal}. The special role of this coupling has already been observed for spin systems, where spin chains have been proven to be controllable with a number of local controls smaller than the number of spins \cite{WangIEEETAC2012}.


\section{Summary and conclusions}\label{sec:conclusion}
We have presented a practical way to test for evolution-operator controllability in order to numerically verify the ability of a given qubit array to implement universal quantum computing. The graph theory-based approach allows for analyzing comparatively large qubit systems for which the evaluation of the Lie rank condition for controllability would be difficult or impossible. Indeed, the computationally expensive calculation of nested commutators of the drift and control Hamiltonians, which is required to construct the system's dynamical Lie algebra, is avoided in graphical methods. On the other hand, in some cases, a graphical controllability test remains inconclusive. We believe that the disadvantage is  small in view of the positive results that can be obtained for systems not amenable to the Lie rank condition.

The key challenge in controllability tests for systems of coupled qubits using graph theory-based methods is due to the tensor product structure of Hilbert space. This leads to graphs that often consist of sets of coupled (resonant) transitions. A similar problem was encountered in the controllability analysis of driven quantum rotors \cite{leibscher2022} which are highly degenerate systems with multiple resonant transitions. In that case, the specific spectral structure allowed for an inductive evaluation of the graphical commutators~\cite{pozzoli2022}. Here, we have made no assumptions on the spectral structure of the system. Instead, we have found and implemented numerically efficient ways to decouple the resonant transitions such as to make them exploitable for the graph test. In particular, our algorithm only determines those graphical commutators which are relevant because they add to the connectivity of the graph. Moreover, we have made use of fact that in qubit arrays, coupled transitions are typically driven by several controls: These coupled transitions become decoupled if the dimension of the subalgebra that they generate when driven by different controls is maximal. Since the sub-algebras are very small compared to the Hilbert space of the complete system, their numerical construction does not pose a challenge.
The remaining limitation to the number of qubits $N_Q$, for which a controllability test can be carried out, is the exponential scaling of the Hilbert space dimension with $N_Q$ because the drift Hamiltonian (including the qubit-qubit couplings) needs to be diagonalized. Typically, the drift is sparse in the logical basis. For the graph test presented here, only a single diagonalization of the (sparse) drift Hamiltonian is necessary to obtain the initial graph. 

We have illustrated the utility of our approach by showing how controllability analysis 
can be used to determine the minimum number of local controls required for universal quantum computing in existing quantum processing units. To this end, we have chosen the five-qubit array \textit{ibmq\_quito} \cite{IBMquito} as specific example, treating the qubit-qubit couplings as fixed. This is justified since, whenever a system with time-independent couplings is controllable, the corresponding system with tunable couplings is controllable as well. The actual \textit{ibmq\_quito} system contains a local control for each of the five qubits. Our analysis has shown that by modifying the couplings between the qubits, the number of local controls necessary for universal quantum computing can be reduced to a single local control. For standard qubit-qubit couplings the minimal number of local controls is two. Analyzing the controllability for different positions of the two local controls reveals that the distance between the local controls is essential for the system to be controllable: The two local controls must be separated by an odd number (one or three) of qubit-qubit couplings in order to provide controllability and allow for universal quantum computing. These results showcase the utility of our graph test for improving the design and scalability of qubit arrays. 

In future work, it will be interesting to gather a deeper understanding of the required control distance, as this might allow for generalizing our findings to larger qubit arrays. Another important open question is the relationship between the minimal number of local controls and the quantum speed limit for universal quantum computing. The quantum speed limit \cite{deffner2017QSL} refers to the minimum time in which a quantum process, such as a quantum gate, can be executed. For a complete universal set of gates, the quantum speed limit can be obtained using quantum optimal control~\cite{goerz2017}. Given a quantum system and a set of controls, the time required for any particular operation will always be equal or longer than the time required for the same operation in the same system with more controls added. Conversely, reducing the number of local controls is likely to increase the time needed to carry out a gate. Thus, reducing the number of local controls has to be balanced with the requirement of sufficiently fast logical operations. The minimal sets of local controls from controllability analysis as suggested here offers a good starting point to find the best balanced set of controls with quantum optimal control.


\section*{Acknowledgments}

We would like to thank David Pohl, Max Werninghaus, Eugenio Pozzoli, Daniel Basilewitsch, Ugo Boscain, and Mario Sigalotti for the helpful discussions. We gratefully acknowledge financial support from the European Union’s Horizon 2020 research and innovation programme under the Marie
Sklodowska-Curie grant agreement Nr. 765267 (QuSCo) and the Einstein Research Foundation (Einstein Research Unit on Near-Term Quantum Devices).



\bibliographystyle{quantum}
\bibliography{main}

\begin{thebibliography}{10}

\bibitem{GlaserEPJD15}
Steffen~J. Glaser, Ugo Boscain, Tommaso Calarco, Christiane~P. Koch, Walter
  K\"ockenberger, Ronnie Kosloff, Ilya Kuprov, Burkard Luy, Sophie Schirmer,
  Thomas Schulte-Herbr\"uggen, D.~Sugny, and Frank~K. Wilhelm.
\newblock ``Training {S}chr\"{o}dinger’s cat: quantum optimal control.
  strategic report on current status, visions and goals for research in
  europe''.
\newblock \href{https://dx.doi.org/10.1140/epjd/e2015-60464-1}{Eur. Phys. J. D
  {\bf 69}, 279}~(2015).

\bibitem{KochEPJQT22}
Christiane~P. Koch, Ugo Boscain, Tommaso Calarco, Gunther Dirr, Stefan Filipp,
  Steffen~J. Glaser, Ronnie Kosloff, Simone Montangero, Thomas
  Schulte-Herbr\"uggen, Dominique Sugny, and Frank~K. Wilhelm.
\newblock ``Quantum optimal control in quantum technologies. strategic report
  on current status, visions and goals for research in europe''.
\newblock \href{https://dx.doi.org/10.1140/epjqt/s40507-022-00138-x}{EPJ
  Quantum Technol. {\bf 9}, 19}~(2022).

\bibitem{dAlessandro2008}
Domenico d'Alessandro.
\newblock ``Introduction to quantum control and dynamics''.
\newblock
  \href{https://dx.doi.org/https://doi.org/10.1201/9781003051268}{Chapman and
  Hall/CRC}. ~(2008).

\bibitem{SchirmerPRA2001}
S.~G. Schirmer, H.~Fu, and A.~I. Solomon.
\newblock ``Complete controllability of quantum systems''.
\newblock \href{https://dx.doi.org/10.1103/PhysRevA.63.063410}{Phys. Rev. A
  {\bf 63}, 063410}~(2001).

\bibitem{FuJPhysA2001}
H~Fu, S~G Schirmer, and A~I Solomon.
\newblock ``Complete controllability of finite-level quantum systems''.
\newblock \href{https://dx.doi.org/10.1088/0305-4470/34/8/313}{Journal of
  Physics A: Mathematical and General {\bf 34}, 1679}~(2001).

\bibitem{AltafiniJMP2002}
Claudio Altafini.
\newblock ``Controllability of quantum mechanical systems by root space
  decomposition of su(n)''.
\newblock \href{https://dx.doi.org/10.1063/1.1467611}{Journal of Mathematical
  Physics {\bf 43}, 2051--2062}~(2002).

\bibitem{chambrion2009}
Thomas Chambrion, Paolo Mason, Mario Sigalotti, and Ugo Boscain.
\newblock ``Controllability of the discrete-spectrum {S}chr{\"o}dinger equation
  driven by an external field''.
\newblock
  \href{https://dx.doi.org/https://doi.org/10.1016/j.anihpc.2008.05.001}{Annales
  de l'Institut Henri Poincar{\'e} C {\bf 26}, 329--349}~(2009).

\bibitem{Boussaid2013}
Nabile Boussa\"{\i}d, Marco Caponigro, and Thomas Chambrion.
\newblock ``Weakly coupled systems in quantum control''.
\newblock \href{https://dx.doi.org/10.1109/TAC.2013.2255948}{IEEE Trans.
  Automat. Control {\bf 58}, 2205--2216}~(2013).

\bibitem{BCCS}
U.~Boscain, M.~Caponigro, T.~Chambrion, and M.~Sigalotti.
\newblock ``A weak spectral condition for the controllability of the bilinear
  {S}chr\"{o}dinger equation with application to the control of a rotating
  planar molecule''.
\newblock \href{https://dx.doi.org/10.1007/s00220-012-1441-z}{Comm. Math. Phys.
  {\bf 311}, 423--455}~(2012).

\bibitem{leibscher2022}
Monika Leibscher, Eugenio Pozzoli, Cristobal P{\'e}rez, Melanie Schnell, Mario
  Sigalotti, Ugo Boscain, and Christiane~P. Koch.
\newblock ``Full quantum control of enantiomer-selective state transfer in
  chiral molecules despite degeneracy''.
\newblock
  \href{https://dx.doi.org/https://doi.org/10.1038/s42005-022-00883-6}{Communications
  Physics {\bf 5}, 1--16}~(2022).

\bibitem{WangIEEETAC2012}
Xiaoting Wang, Peter Pemberton-Ross, and Sophie~G. Schirmer.
\newblock ``Symmetry and subspace controllability for spin networks with a
  single-node control''.
\newblock \href{https://dx.doi.org/10.1109/TAC.2012.2202057}{IEEE Transactions
  on Automatic Control {\bf 57}, 1945--1956}~(2012).

\bibitem{WangPRA2016}
Xiaoting Wang, Daniel Burgarth, and S.~Schirmer.
\newblock ``Subspace controllability of spin-$\frac{1}{2}$ chains with
  symmetries''.
\newblock \href{https://dx.doi.org/10.1103/PhysRevA.94.052319}{Phys. Rev. A
  {\bf 94}, 052319}~(2016).

\bibitem{ChenPRA2020}
Jiahui Chen, Yehao Zhou, Ji~Bian, Jun Li, and Xinhua Peng.
\newblock ``Subspace controllability of symmetric spin networks''.
\newblock \href{https://dx.doi.org/10.1103/PhysRevA.102.032602}{Phys. Rev. A
  {\bf 102}, 032602}~(2020).

\bibitem{albertini2021subspace}
Francesca Albertini and Domenico D’Alessandro.
\newblock ``Subspace controllability of multi-partite spin networks''.
\newblock
  \href{https://dx.doi.org/https://doi.org/10.1016/j.sysconle.2021.104913}{Systems
  \& Control Letters {\bf 151}, 104913}~(2021).

\bibitem{AlbertiniLinAlg2002}
Francesca Albertini and Domenico D'Alessandro.
\newblock ``The {L}ie algebra structure and controllability of spin systems''.
\newblock
  \href{https://dx.doi.org/https://doi.org/10.1016/S0024-3795(02)00290-2}{Linear
  Algebra and its Applications {\bf 350}, 213--235}~(2002).

\bibitem{boscain2014multi}
Ugo Boscain, Marco Caponigro, and Mario Sigalotti.
\newblock ``Multi-input {S}chr{\"o}dinger equation: controllability, tracking,
  and application to the quantum angular momentum''.
\newblock
  \href{https://dx.doi.org/https://doi.org/10.1016/j.jde.2014.02.004}{Journal
  of Differential Equations {\bf 256}, 3524--3551}~(2014).

\bibitem{godsil2010graph}
Chris Godsil and Simone Severini.
\newblock ``Control by quantum dynamics on graphs''.
\newblock
  \href{https://dx.doi.org/https://doi.org/10.1103/PhysRevA.81.052316}{Physical
  Review A {\bf 81}, 052316}~(2010).

\bibitem{GoklerPRL17}
Can Gokler, Seth Lloyd, Peter Shor, and Kevin Thompson.
\newblock ``Efficiently controllable graphs''.
\newblock \href{https://dx.doi.org/10.1103/PhysRevLett.118.260501}{Phys. Rev.
  Lett. {\bf 118}, 260501}~(2017).

\bibitem{pozzoli2022}
Eugenio Pozzoli, Monika Leibscher, Mario Sigalotti, Ugo Boscain, and
  Christiane~P. Koch.
\newblock ``{L}ie algebra for rotational subsystems of a driven asymmetric
  top''.
\newblock \href{https://dx.doi.org/https://doi.org/10.1088/1751-8121/ac631d}{J.
  Phys. A: Math. Theor. {\bf 55}, 215301}~(2022).

\bibitem{figgatt2019ion}
Caroline Figgatt, Aaron Ostrander, Norbert~M Linke, Kevin~A Landsman, Daiwei
  Zhu, Dmitri Maslov, and Christopher Monroe.
\newblock ``Parallel entangling operations on a universal ion-trap quantum
  computer''.
\newblock
  \href{https://dx.doi.org/https://doi.org/10.1038/s41586-019-1427-5}{Nature
  {\bf 572}, 368--372}~(2019).

\bibitem{krantz2019quantum}
Philip Krantz, Morten Kjaergaard, Fei Yan, Terry~P Orlando, Simon Gustavsson,
  and William~D Oliver.
\newblock ``A quantum engineer's guide to superconducting qubits''.
\newblock \href{https://dx.doi.org/https://doi.org/10.1063/1.5089550}{Applied
  Physics Reviews {\bf 6}, 021318}~(2019).

\bibitem{schirmer2002identification}
SG~Schirmer, ICH Pullen, and AI~Solomon.
\newblock ``Identification of dynamical {L}ie algebras for finite-level quantum
  control systems''.
\newblock
  \href{https://dx.doi.org/https://doi.org/10.1088/0305-4470/35/9/319}{Journal
  of Physics A: Mathematical and General {\bf 35}, 2327}~(2002).

\bibitem{boscain2021classical}
Ugo Boscain, Eugenio Pozzoli, and Mario Sigalotti.
\newblock ``Classical and quantum controllability of a rotating symmetric
  molecule''.
\newblock \href{https://dx.doi.org/https://doi.org/10.1137/20M1311442}{SIAM
  Journal on Control and Optimization {\bf 59}, 156--184}~(2021).

\bibitem{IBMquito}
IBM.
\newblock ``{IBM} quantum''.
\newblock
  url:~\href{https://quantum-computing.ibm.com/services}{quantum-computing.ibm.com/services}.

\bibitem{roth2017analysis}
Marco Roth, Marc Ganzhorn, Nikolaj Moll, Stefan Filipp, Gian Salis, and
  Sebastian Schmidt.
\newblock ``Analysis of a parametrically driven exchange-type gate and a
  two-photon excitation gate between superconducting qubits''.
\newblock
  \href{https://dx.doi.org/https://doi.org/10.1103/PhysRevA.96.062323}{Physical
  Review A {\bf 96}, 062323}~(2017).

\bibitem{deffner2017QSL}
Sebastian Deffner and Steve Campbell.
\newblock ``Quantum speed limits: from {H}eisenberg’s uncertainty principle
  to optimal quantum control''.
\newblock
  \href{https://dx.doi.org/https://doi.org/10.1088/1751-8121/aa86c6}{Journal of
  Physics A: Mathematical and Theoretical {\bf 50}, 453001}~(2017).

\bibitem{goerz2017}
Michael~H. Goerz, Felix Motzoi, K.~Birgitta Whaley, and Christiane~P. Koch.
\newblock ``Charting the circuit-{QED} design landscape using optimal control
  theory''.
\newblock \href{https://dx.doi.org/10.1038/s41534-017-0036-0}{npj Quantum Inf.
  {\bf 3}, 37}~(2017).

\end{thebibliography}

\end{document}